\documentclass[fleqn,twoside]{article}
\usepackage{gc}

\usepackage{latexsym,amsmath}
\renewcommand{\theequation}{\thesection\arabic{equation}}
\newcommand{\be}{\begin{equation}}
\newcommand{\ee}{\end{equation}}
\newcommand{\ben}{\begin{enumerate}}
\newcommand{\een}{\end{enumerate}}
\newcommand{\beqs}{\begin{eqnarray*}}
\newcommand{\eeqs}{\end{eqnarray*}}

\newcommand{\la}{\label}
\newcommand{\fp}[2]{\frac{\partial #1}{\partial #2}}

\newcommand{\bmin}{\begin{minipage}}
\newcommand{\emin}{\end{minipage}}

\newcommand{\llq}{\lq\lq}
\newcommand{\rrq}{\rq\rq}

\newcommand{\bt}{\beta}

\newcommand{\lam}{\lambda}

\newcommand{\g}{\gamma}

\newcommand{\pdr}{\partial}

\newenvironment{definicion}[1]{\begin{flushleft} $\bullet$ {\bf \Large #1 \hfill}}{\end{flushleft}}
\newcommand{\bedef}{\begin{definicion}}
\newcommand{\eedef}{\end{definicion}}

\heads{J. Quiroga }
      {Spontaneous Symmetry Breaking in the Brane-world}

\begin{document}
\twocolumn[ \Arthead{6}{2000}{4 (24)}{1}{10}

\Title{Spontaneous Symmetry Breaking in              \yy
       the Brane-World}

\Author{John Quiroga H.\foom 1 \foom 2}
       {Department of Physics , Universidad Tecnologica de
Pereira, Colombia}{and Tomsk State Pedagogical University, Tomsk
634041, Russia}

\Abstract{A simplified  Randall-Sundrum-like model in 6 dimensions
is discussed. The extra two dimensions correspond to the cone. The
effective four-dimensional scalar self-interacting theory is
studied at one-loop level. The contributions due to 6-dimensional
parameters in four-dimensional beta-functions appear. Using such
beta-functions the one-loop effective potential is calculated. The
possibility of spontaneous symmetry breaking due to extra
dimensions is demonstrated.}

]

\email 1 {jquiroga@tspu.edu.ru} \email 2{jquiroga@utp.edu.ru}
\section{Introduction}
The spontaneous symmetry breaking  effect in an external
gravitational field has a lot of very interesting characteristics.
In particular it is important to remark that the interaction with
the external gravitational field may lead to the spontaneous
symmetry breaking (see papers \cite{Janson,Grib}). It is possible
to give masses for gauge fields without the introduction of a \llq
negative square\rrq \,  mass in the scalar sector. One way of
doing this is for example to introduce the term $\xi R$, which
gives the non-minimal coupling between the scalar and the
gravitational fields (\cite{buch,Janson,Grib}).

The influence of the quantum corrections on the spontaneous
symmetry breaking is easier to study on the basis of effective
potential (EP). Approximate expressions for EP in an external
gravitational field of the special form  have been obtained in
papers (see for example \cite{OSDB,LR,VLR,DS}), for the theory of
the scalar field with interaction $\lam \phi^4$, scalar
electrodynamics and gauge theory with scalars.

In this paper we consider the theory $\lam \phi^4$ with mass in a
six-dimensional space-time. For this theory we are interesting in
the influence of the extra two dimensions on the spontaneous
symmetry breaking in the effective four-dimensional theory.

An interesting way to do this is to consider the Randall-Sundrum
brane-world models \cite{RS}, which at the present time already
has attracted a great deal of interest in particle physics and
phenomenology because of the possibility to resolve the hierarchy
problem in a quite natural way.

In a typical approach to modern quantum high energy theory it is
necessary to consider quantum fields on a higher dimensional
manifold (the bulk) in the presence of extended defects (the
boundaries). On the bulk manifold, as well as on the brane, there
exist divergences which result in the running of coupling
constants in the standard way. However, it has been known for some
time that for spacetimes with boundaries there are not only the
usual volume coupling constants but also surface ones
\cite{surface}. It is known that they influence each other; for
example volume interactions are reflected in surface terms, etc
\cite{tsoupros}. But the situation could be complicated in some
brane-world models. For example one may wonder how the running of
the bulk couplings influences the running on the brane and
vice-versa. It has been demonstrated in ref.\cite{milton} that
bulk contribution may completely change the standard running
behaviour of brane couplings as the effect of bulk couplings.
Moreover one may wonder how this influences the spontaneous
symmetry breaking. This is the purpose of this work to study the
spontaneous symmetry breaking in the model of ref.\cite{milton}.

\section{Description of the Model}

Our model is given as a massive Euclidean self-interacting scalar
in a 6-dimensional space with a conical singularity, due to the
presence of 3-brane. The metric is chosen to be \cite{milton}
\begin{equation}
ds_6^2=dr^2+r^2d\theta^2+ds^2_0\,, \label{111}
\end{equation}
where $ds^2_0$ is the 4-dimensional flat metric, the brane is
located at $r=0$, and $\theta$ has a period $\beta$, $\beta $
being the deficit angle of the cone. When $\beta=2 \pi/N$, $N$ a
positive integer, one is dealing with a less singular manifold,
namely an orbifold, while for $N=1$, $\beta= 2\pi$, one  has the
smooth 2-dimensional plane. The action  reads
\begin{equation}
S= \int d^6x\, \sqrt{g} \left[ \phi  \frac{1}{2}(-\Box_6+m^2  )
\phi+ V(\phi) \right]+\int d^4x\, W(\phi)\,, \label{2}
\end{equation}
where $V(\phi)=\frac{g_4}{4!}\phi^4+\dots$ denotes a series of
scalar  bulk couplings. We also introduce a ``surface"  term which
depends on surface scalar couplings
\begin{equation}
W(\phi)=\left[\lambda_0+\frac{\lambda_2}{2}
\phi^2+\frac{\lambda_4}{4!} \phi^4+\dots\right]\,,
\end{equation}
namely it may contain a brane tension $\lambda_0$, a brane mass
$\lambda_2$, a $\phi^4$ coupling $\lambda_4$, as well as higher
terms. As it has been demonstrated in ref.\cite{milton}, these
surface terms are necessary because we are dealing with a manifold
with a conical singularity. We also assume that the brane is not
dynamical, namely we are dealing with a rigid brane, and therefore
we  neglect the brane kinetic term. This finishes our description
of the model, for more details one may consult in \cite{milton}.

\section{One-loop effective potential}

The one-loop correction is determined by the  total one-loop
fluctuation operator, which  reads
\begin{equation}
L_6 = -\Box_\beta -\Box_4+M^2+W''(\Phi)\delta^{(2)}(x)\,,
\label{3}
\end{equation}
where $\Box_\beta$ is the 2-dimensional Laplacian on the cone, $
\Phi $ is the background field and $M^2=m^2+V''(\Phi)$ is an
effective mass.

We shall make use of zeta-function regularization and related
heat-kernel techniques (see, for example Ref.~\cite{elib}). Within
the one-loop approximation, one needs to evaluate the
zeta-function at zero, namely $\zeta(0|L_6)$, since this quantity
gives rise to the one-loop divergences and governs the one-loop
beta functions. There are also contributions due to the conical
singularity and the brane delta-function contribution, which gives
additive contributions to $\zeta(0|L_6)$, which have been
diagrammatically evaluated in Ref.~\cite{wise}. We will take the
final results for the calculation of four-dimensional effective
beta-functions in refs.\cite{wise,milton}.

In our theory we are taking in account that four-dimensional
curvature $R=0$ and $\phi = const$. For this case theory maybe
considered as one-loop renormalizable and the technique of
refs.\cite{OSDB,LR} to find the one-loop effective potential maybe
used. The one-loop RGE equation for potential is

\be\la{RGE} DL_{eff}=0 \ee

here $D\ \hbox{and}\ L_{eff}$ are defined in \cite{OSDB} as

\begin{eqnarray}
\!\!\!\!\!\!\!\!\!D&=&\mu \fp{ }{\mu}+\bt_\lam \fp{ }{\lam}+\g
\fp{ }{m^2}+\g_\phi \fp{ }{\phi}\\
\!\!\!\!\!\! L_{eff}&=&-a(\lam_4+E_{\lam_4}t)\phi^4-b^2(\lam_2+E_m
t)\phi^2 \quad \la{efflag} \end{eqnarray}

In our situation $E_{\lam_4}=\bt_{\lam_4}\ \hbox{and}\ E_\g = \g$,
where the $\bt_{\lam_4}\ \hbox{and}\ \g$ functions derived in
\cite{milton} have the form

\begin{eqnarray}
\g &=& \frac{\lambda_2^2}{\pi}+ \frac{m^2
g_4}{128\pi^2}-\frac{m^4\lambda_4}{64\pi^3}\\
\bt_{\lam_4}&=&\frac{4\lambda_2\lambda_4}{\pi}+ \frac{3
g_4^2}{128\pi^2}-\frac{m^4\lambda_6}{64\pi^3}\,\label{betafun}
\end{eqnarray}

We add to equations (\ref{efflag}) the renormalization conditions

\begin{equation}
\frac{\partial^4 L_{eff}}{\partial
\phi^4}\Bigg\vert_{\phi=\phi_{\lam_4}}=-4!a \lam_4\, ,\qquad
\frac{\partial^2 L_{eff}}{\partial
\phi^2}\Bigg\vert_{\phi=\phi_m}=-2b\lam_2\, \la{Rcond}
\end{equation}

Using these conditions in (\ref{efflag}) one finally obtains for
$L_{eff}$ \cite{OSDB,LR}

\be \la{Efflag} L_{eff}=
-\frac{1}{4!}[\lam_4+\frac{1}{2}\bt_{\lam_4}\ln\frac{\tau}{m_{\lam_4}^2}]\phi^4-\frac{1}{2}[\lam_2+\frac{1}{2}\g\ln\frac{\tau}{m_m^2}]\phi^2
\ee

Here

\be\la{param} \tau=\lam_2+\frac{\lam_4}{2}\phi^2 \, ,\quad
m^2_{\lam_4}=\lam_2+\frac{\lam_4}{2}\phi^2_{\lam_4}\, ,\quad
m^2_m=\lam_2+\frac{\lam_4}{2}\phi^2_m \ee This finishes the
calculation of one-loop effective potential.

\section{Spontaneous symmetry breaking}

In a flat space with $\phi=const$ the existence of an absolute
minimum of the EP where EP takes negative value may lead to
spontaneous symmetry breaking. One can find the critical value of
$\phi$ from the conditions

\be\la{ssb} \fp{V_{eff}}{\phi}=0\, ,\qquad \frac{\pdr^2
V_{eff}}{\pdr\phi^2}>0 \ee

The EP is given by (\ref{Efflag}) and it has the form

\be \la{EP}
V_{eff}=\frac{1}{4!}[\lam_4+\frac{1}{2}\bt_{\lam_4}\ln\frac{\tau}{m_{\lam_4}^2}]\phi^4+\frac{1}{2}[\lam_2+\frac{1}{2}\g\ln\frac{\tau}{m_m^2}]\phi^2
\ee

In order to study the spontaneous symmetry breaking one may
consider several cases for the relations between the parameters
$\lam_2\, ,\quad \lam_4\, ,\quad \lam_6\, ,\quad g_4$, etc and
obtain an approximation for the EP. Using the conditions
(\ref{ssb}) it is possible to see if there is spontaneous symmetry
breaking for each of the cases. As an example, let us consider
that $\lam_2 \ll \frac{\lam_4}{2}\phi^2$ for all of the values for
$\phi$ . It is not difficult to see that in such case it follows
that

\be \la{EP1} V_{eff}=\frac{1}{4!}\lam_4\phi^4+\frac{1}{2}\lam_2
\phi^2 +
[\frac{1}{48}\bt_{\lam_4}\ln\frac{\phi^2}{\phi^2_{\lam_4}}\,
\phi^4 + \frac{1}{4}\g\ln\frac{\phi^2}{\phi^2_m}\, \phi^2] \ee

Now for simplicity we will consider that $\phi_{\lam_4}=\phi_m =
\phi_o$. Introducing the dimensionless variable
$x=\frac{\phi^2}{\phi^2_o}$ we may write the EP in the following
form

\be \la{EPadim} V_{eff}=\frac{1}{4!}[\lam_4 +
\frac{1}{2}\bt_{\lam_4}\ln x]\phi_o^4 x^2 + \frac{1}{2}[\lam_2 +
\frac{1}{2}\g \ln x]\phi_o^2 x \, ,\ee

Finally for calculation it is better to write this expression as

\be \la{EPadim1} \frac{V_{eff}}{\phi_o^4}=\frac{1}{4!}\lam_4 x^2 +
\frac{1}{2} \frac{\lam_2}{\phi_o^2}x + \Big (
\frac{1}{48}\bt_{\lam_4}x\ln x + \frac{\g}{4\phi_o^2}\ln x \Big )
x \ee

The first derivative for (\ref{EPadim1}) is

\be \la{1der}\begin{array}{ll} \fp{ }{x}\Big (
\frac{V_{eff}}{\phi_o^2}\Big ) = (\frac{1}{6}\lam_4 +
\frac{1}{48}\bt_{\lam_4})x + \\ \\(\frac{1}{24}\bt_{\lam_4}x +
\frac{\g}{4\phi_o^2})\ln x +
(\frac{1}{2}\frac{\lam_2}{\phi_o^2}+\frac{\g}{4\phi_o^2})
\end{array}\ee

Now let us write the first expression in (\ref{ssb}) when $x$
takes values around 1. Thus doing $x=1+z$, where $z \ll 1$, we
obtain the equation

\be\la{eqcuad} \begin{array}{ll}\frac{1}{24}\bt_{\lam_4}z^2 +
(\frac{1}{72}\bt_{\lam_4}+\frac{\g}{4}\phi_o^2 +
\frac{1}{6\lam_4})z +\\  \\ (\frac{1}{48}\bt_{\lam_4} +
\frac{1}{6}\lam_4 + \frac{1}{2}\frac{\lam_2}{\phi_o^2} +
\frac{\g}{4\phi_o^2}) = 0  \end{array}\ee

Before solving this equation it is recommended to take the second
derivative of (\ref{EPadim1}), in order to verify then the second
condition for the spontaneous symmetry breaking. Thus we have

\be \la{2der} \frac{\pdr^2}{\pdr x^2}\Big (
\frac{V_{eff}}{\phi_o^2}\Big ) = \frac{1}{6}\lam_4 +
\frac{1}{24}\bt_{\lam_4} + \frac{1}{24}\bt_{\lam_4}\ln x +
\frac{\g}{4\phi_o^2}\frac{1}{x} + \frac{1}{48}\bt_{\lam_4} \ee

Solving for $z$ the equation (\ref{eqcuad}) we will use the
approximation

$$\bt_{\lam_4} \approx -\frac{m^4 \lam_6}{64\pi^3}\, , \quad \g
\approx \frac{m^2 g_4}{128\pi^2}\,  \quad \hbox{and}\quad m^2g_4
\gg m^4 \lam_6 $$

This approximation leads to two possible solutions  for $z$ in
(\ref{eqcuad}).

The first solution is $z=0$ and this value gives us  $x=1$. It is
not difficult to see that in this case the second derivative
(\ref{2der}) satisfies the second condition for the spontaneous
symmetry breaking.

The second solution
$$z=\frac{3\pi}{2\phi_o^2}\frac{m^2g_4}{m^4\lam_4}$$
gives us a big value of $x$ and it is easy to see that the second
derivative will be negative, so with this solution there is no
spontaneous symmetry breaking.

Now let us consider that $x \approx 0$. The equation for the first
derivative is written now in the form

\be \la{1der2c} \frac{\g}{4\phi_o^2}\ln x +
(\frac{1}{2}\frac{\lam_2}{\phi_o^2}+\frac{\g}{4\phi_o^2})=0 \ee

The solution for this equation can be written as

\be \la{x2c} x=e^{-(1+\frac{2\lam_2}{\g})}  \ee

This value of $x$ with the approximations that were used before
leads to spontaneous symmetry breaking too.

Finally we may say that using the same approximation for the
situation when $x \gg 1$, it is easy to find an absolute minimum
of the EP, which indicates that there is spontaneous symmetry
breaking in this case too.

Thus, we demonstrated that there maybe spontaneous symmetry
breaking induced by extra two dimensions (in our case chosen as a
cone). It occurs mainly due to parameters of higher dimensional
theory. That indicates that extra dimensions in brane-worlds may
play an important role in phenomena like Higgs effect and
spontaneous symmetry breaking. In curved four-dimensional space
one expects that there maybe phase transitions induced by extra
dimensions what could find various applications in early universe
cosmology.

\Acknow {I am grateful to Prof.S.D. Odintsov for formulation the
problem and numerous helpful discussions. I thank Prof.P.M. Lavrov
for very useful discussions. This research was supported in part
by Professorship and Fellowship from the Universidad Tecnologica
de Pereira, Colombia.}

\end{document}